\documentclass{ws-procs961x669}            
\begin{document}
\title{Testing extended theories of gravity with GRBs}

\author{L. Mastrototaro, G. Lambiase}

\address{Dipartimento di Fisica ``E.R Caianiello'', Università degli Studi di Salerno,\\ Via Giovanni Paolo II, 132 - 84084 Fisciano (SA), Italy
\\
\vspace{0mm}
Istituto Nazionale di Fisica Nucleare - Gruppo Collegato di Salerno,\\ Via Giovanni Paolo II, 132 - 84084 Fisciano (SA), Italy.
$^*$E-mail: lmastrototaro@unisa.it}

\begin{abstract}
We present our studies on the
neutrino pairs annihilation into electron-positron pairs ($\nu{\bar \nu}\to e^-e^+$) near the surface of a neutron star in the framework of extended theories of gravity. The latter modifies the maximum energy deposition rate near to the photonsphere and it might be several orders of magnitude greater than that computed in the framework of General Relativity. These results provide a rising in the Gamma-Ray Bursts energy emitted from a close binary neutron star system and might be a fingerprint of modified theories of gravity, changing our view of astrophysical phenomena.
\end{abstract}

\keywords{Extended theories of gravity, GRB, Neutrino energy deposition, Black Holes, Neutron Stars}

\bodymatter

\section{Introduction}
\label{Introduction}
General Relativity (GR) is without any doubt the best theory of gravitational interaction. Its predictions have been tested with high accuracy on scales of the solar system (for example, the precession of the Mercury perihelion, the photons deviation and the gravitational leasing effect), on astrophysical scales (the gravitational waves), and cosmological scales (the cosmic microwave background radiation (CMBR)  and the formation of primordial light elements (Big Bang Nucleosynthesis)). Despite these results, there are still open questions that make GR incomplete. The latter arise at short distances and small time scales (black hole and cosmological singularities, respectively), or at large distance scales, the rotational curve of the galaxies and the observed accelerated phase of the present Universe, for which any predictability is lost. 

To solve these issues, deviations from the GR (hence from the Hilbert-Einstein action on which GR is based) are needed, and new ingredients, such as dark matter and dark energy, are required for fitting the present picture of our Universe \cite{riess,riess1,riess2,riess3,riess4,riess5}. Indeed, in the last years, several {\it alternative} or {\it modified} theories of gravity have been proposed, which try to answer all the opened questions of GR and the Cosmological Standard Model. To give an example, higher-order curvature invariants than the simple Ricci scalar $R$, allow getting inflationary behaviour, removing the primordial singularity, as well as explaining the flatness and horizon problems \cite{starobinski,starobinski1}( for further applications, see Refs. \cite{cosmo2,cosmo3,cosmo4,cosmo5,cosmo6,cosmo7,cosmo8,cosmo9,cosmo10,cosmo11,cosmo12,
cosmo13,cosmo14,cosmo15,cosmo17,cosmo18,cosmo19,cosmo20,cosmo21,cosmo22,cosmo23,cosmo24,cosmo25,Buoninfante:2021qrv,Capolupo:2021blb,Bittencourt:2020lgu,Lambiase:2020vul,Bernal:2020ywq,Tino:2020nla}). On the other hand, one can tend to preserve the GR and the Hilbert-Einstein action, adding only some two unidentified components: dark matter and dark energy.
These two different approaches try to solve the same problems but at the moment there is not a final solution for that (modified gravity theories do not manage to solve all the problems while dark matters elements are still missing). One of the consequences of these approaches is that the metric tensor $g_{\mu\nu}$ describing the gravitational field generated by a massive source gets modified with respect to GR metric, in particular, the Schwarzschild or Kerr metric. The latter are recovered in the limit in which the parameters characterizing some specific theory of gravity beyond GR are set to zero.

In this proceeding, we highlight the differences between GR and extended theories of gravity arising from the mechanism of generation of Gamma-Ray Bursts (GRBs). We focus, in particular, on GRBs powered by neutrino annihilation processes $\nu\Bar{\nu}\rightarrow e^+e^-$ \footnote{Indeed the state of the art of NS-NS merger simulations is special relativistic, axisymmetric hydrodynamic simulations for the final black hole-torus systems~\cite{1999A&A...344..573R,Popham_1999,Di_Matteo_2002,Fujibayashi_2017,Just:2015dba,PhysRevD.98.063007} from which emerge that neutrino-pair annihilation in ordinary GR models seems to be not efficient enough to power GRBs and the Blandford-Znajek process is currently considered a more promising mechanism for launching. In extended theories of gravity, the situation can be different as we will show in this proceeding.}. 

The neutrinos annihilation process is relevant in many astrophysical frameworks: in the stellar envelope, as well as on the delay shock mechanism into the Type II Supernova (at late time, from the hot proto-neutron star, the energy is deposited into the supernova envelope via neutrino pair annihilation and neutrino-lepton scattering). These processes augment the neutrino heating of the envelope generating a successful supernova explosion \cite{Salmonson:1999es}. Moreover, the neutrino annihilation process has been also proposed as the mechanism that power GRB in a binary neutron star system, which is the topic of this work. Simulations and analytical estimations performed within GR show that the mechanism is not sufficient to generate the required energy for explaining short GRB. Such a conclusion changes, as we will show, if the gravitational background is described by modified theories.

\section{Neutrino energy deposition formulation}
\label{Formulation}
In this section we recall the main features to treat the energy deposition in curved spacetimes  \cite{Prasanna:2001ie,Lambiase:2020iul}.  Previous calculations of the $\nu {\bar \nu}\to e^- e^+$ reaction in the vicinity of a neutron star have been first based on Newtonian gravity \cite{Co86,Co87}, then the effect of gravity has been incorporated for static stars \cite{Salmonson:1999es,Salmonson:2001tz}, and then extended to
rotating stars \cite{Prasanna:2001ie,Bhattacharyya:2009nm}. We consider the spacetime around a black hole described by the following diagonal metric 
  \begin{equation}
g_{\mu\nu}=\text{diag}\left(-f(r), h(r), r^2, r^2 \sin^2 \theta\right)\,,
\label{metric}
\end{equation}


The energy deposition per unit time and per volume is given by (considering $c=\hbar =1$)  \cite{Goodman:1986we} 
\begin{equation}\label{qdotgeneral}
\begin{split}
\dot{q}(r) = & \iint f_\nu({\bf p}_\nu,r)
f_{\overline{\nu}}({\bf p}_{\overline{\nu}},r) \times \\&
 \times \Big[\sigma |{\bf v}_\nu - {\bf v}_{\overline{\nu}} |  \varepsilon_\nu
\varepsilon_{\overline{\nu}} \Big] 
\frac{ \varepsilon_\nu + \varepsilon_{\overline{\nu}}}
{
 \varepsilon_\nu \varepsilon_{\overline{\nu}} } d^3{\bf p}_\nu d^3{\bf p}_{\overline{\nu}}\,,
\end{split}
\end{equation}
where $f_{\nu,{\overline{\nu}}}$ are the neutrino number densities in phase space, ${\bf v}_\nu$ the neutrino velocity, and $\sigma$ is the rest frame cross section.  
Since the term $\sigma |{\bf v}_\nu - {\bf v}_{\overline{\nu}} | \varepsilon_\nu
\varepsilon_{\overline{\nu}}$ is Lorentz invariant, it can be calculated in the center-of-mass frame, and turns out to be
\begin{equation}
\sigma |{\bf v}_\nu - {\bf v}_{\overline{\nu}} | \varepsilon_\nu
\varepsilon_{\overline{\nu}}  = \frac{D G^2_F}{3\pi} (\varepsilon_\nu
\varepsilon_{\overline{\nu}} - {\bf p}_\nu \cdot
{\bf p}_{\overline{\nu}} c^2 )^2\,,
\end{equation}
with $G_F=5.29 \times 10^{-44}$ cm$^2$ MeV$^{-2}$ the Fermi constant,
\begin{equation}
D=1\pm4\sin^2\theta_W+8\sin^4\theta_W, 
\end{equation}
$\sin^2\theta_W=0.23$ the Weinberg angle, and the plus sign for electron neutrinos and antineutrinos while the minus sign for muon and tau type. $T(r)$ is the temperature measured by the local observer and $\Theta(r)$ is the angular integration factor. At these energies, the mass of the electrons can be neglected and it is possible to obtain that the general expression of the rate per unit time and unit volume of the $\nu\bar\nu\rightarrow e^+e^-$ process \cite{Salmonson:1999es} 
\begin{equation}\label{qpunto}
\dot{q}=\frac{7DG_F^2\pi^3\xi(5)}{2}[k T(r)]^9\Theta(r) \,.
\end{equation}
 The evaluation of $T(r)$ and $\Theta(r)$ account for the gravitational redshift and path bending. To write the latter in terms of observed luminosity $L_\infty$ one has to has to consider that temperature, like energy, varies linearly with red-shift and following the procedure of Ref. \cite{Salmonson:1999es}, one finds that:
\begin{align}
    \Theta(r)&=\frac{2\pi^3}{3}(1-x)^4(x^2+4x+5) \,, \\
    T(r)&=\frac{\sqrt{f(R)}}{\sqrt{f(r)}}T(R) \,, \\
    L_{\infty}&=f(R)L(R) \,, \\
    L(R) & = L_\nu + L_{\overline{\nu}} = \frac{7}{4} \, a \pi R^2 T^4(R) \,.
\end{align}
Here $R$ is the neutrinosphere radius (the spherical surface where the stellar material is transparent to neutrinos and from which neutrinos are emitted freely), $L(R)$ is the neutrino luminosity, $a$ the radiation constant, $x=\sin^2\theta_r$, $\theta_r$ is the angle between the trajectory and the tangent velocity in terms of local radial and longitudinal velocities~\cite{Prasanna:2001ie} for which one can obtain that \cite{Lambiase:2020iul}
\begin{equation}\label{cosr}
\cos\theta_r=\frac{R}{r}\sqrt{\frac{f(r)}{f(R)}}.
\end{equation}
This relation comes from the fact that the impact parameter $b$ is constant on all the trajectory and is related to $\cos\theta_r$ by the relation
\begin{equation}\label{bSchw}
    b=\left(\frac{f(r)}{r\cos\theta_r}\right)^{-1} \,\ .
\end{equation}
Moreover, photosphere radius $R_{ph}$ exists below which a massless particle can not be emitted tangent to the stellar surface. The present discussion is therefore restricted to $R>R_{ph}$. The neutrino emission properties hence mainly depend on the geometry of spacetime. From the equation of the velocities 
\[
\dot{r}^2 = \left(E\dot{t}-L\dot{\phi}\right)f(r)\,, \quad
\dot{\phi}=\frac{L}{r^2}, \quad
\dot{t} =-\frac{E}{f(r)},
 \]
where $E$ and $L$ are the energy and angular momentum at the infinity, one gets the effective potential $V_{eff}$, such that the photonsphere radius follows from the condition $\frac{\partial V_{\mathrm{eff}}}{\partial r}=0$. The circular orbit is derived by imposing $\dot{r}^2=0$.
We note en passant that these results reduce to ones derived in the case of the Schwarzschild geometry, $R_{ph}=3M$, as calculated in \cite{Salmonson:1999es}.

The integration of $\dot{q}$ from $R$ to infinity gives the total amount of local energy deposited by the neutrino annihilation process (for a single neutrino flavour) for time units
\begin{equation}\label{Qdot}
    \dot{Q}= 4\pi \int_R^{\infty}  dr \frac{r^2}{\sqrt{f(r)}}\, \dot{q} \,.
\end{equation}
According to \cite{Salmonson:1999es},  the total energy deposition from the neutrinosphere radius $R_6$ to infinity (for a symmetric spherical star that emits neutrinos from a spherical neutrinosphere) is given by
\begin{equation}
\dot{Q}_{51}=1.09\times 10^{-5}\mathcal{F}\left(\frac{M}{R}\right)DL_{51}^{9/4}R_6^{-3/2} \,.
\label{contoQ}
\end{equation}
Here $\dot{Q}_{51}$ is expressed in units of $10^{51}~\mathrm{erg~s^{-1}}$, $L_{51}$ is neutrino luminosity in units of $10^{51}~\mathrm{erg~s^{-1}}$, $D=1\pm4\sin^2\theta_W+8\sin^4\theta_W$, $\sin^2\theta_W=0.23$ and the plus sign is for electron neutrinos and antineutrinos while the minus sign is for muon and tau type, $R_6$ is the radius in units of $10~\mathrm{km}$ and, for a generic diagonal metric of the form $g_{\mu\nu}=\left(g_{00},-1/g_{00},-r^2,-r^2\sin^2\theta\right)$, the function $\mathcal{F}\left(\frac{M}{R}\right)$ is given by 
\begin{equation}
\mathcal{F}\left(\frac{M}{R}\right)=3g_{00}(R)^{9/4}\int_1^{\mathrm{R_{ch}}}(x-1)^4(x^2+4x+5)\frac{y^2g_{11}(yR)^{1/2}dy}{g_{00}(yR)^{9/2}} \,\ .
\label{f}
\end{equation}
In the Newtonian limit one gets $\mathcal{F}(0)=1$ so that it is convenient, for our analysis, to consider the ratio $\dot{Q}_{\mathrm{GR}}/\dot{Q}_{\mathrm{Newt}}=\mathcal{F}(M/R)$. Usually, almost all energy is carried out by electron neutrino thus we can approximate Eq.~(\ref{contoQ}) considering $D=1.23$.

Equations (\ref{contoQ}) and (\ref{f}) allow obtaining the deposited energy of neutrinos and the energy that can be emitted to powering GRB. 
In the next Section, we will study the ratio (\ref{f}) for astrophysical objects in various modified gravity theories.

\section{Neutrino deposition in modified gravity}
\label{Results}
\begin{figure}[h!]
\centering
\includegraphics[scale=0.6]{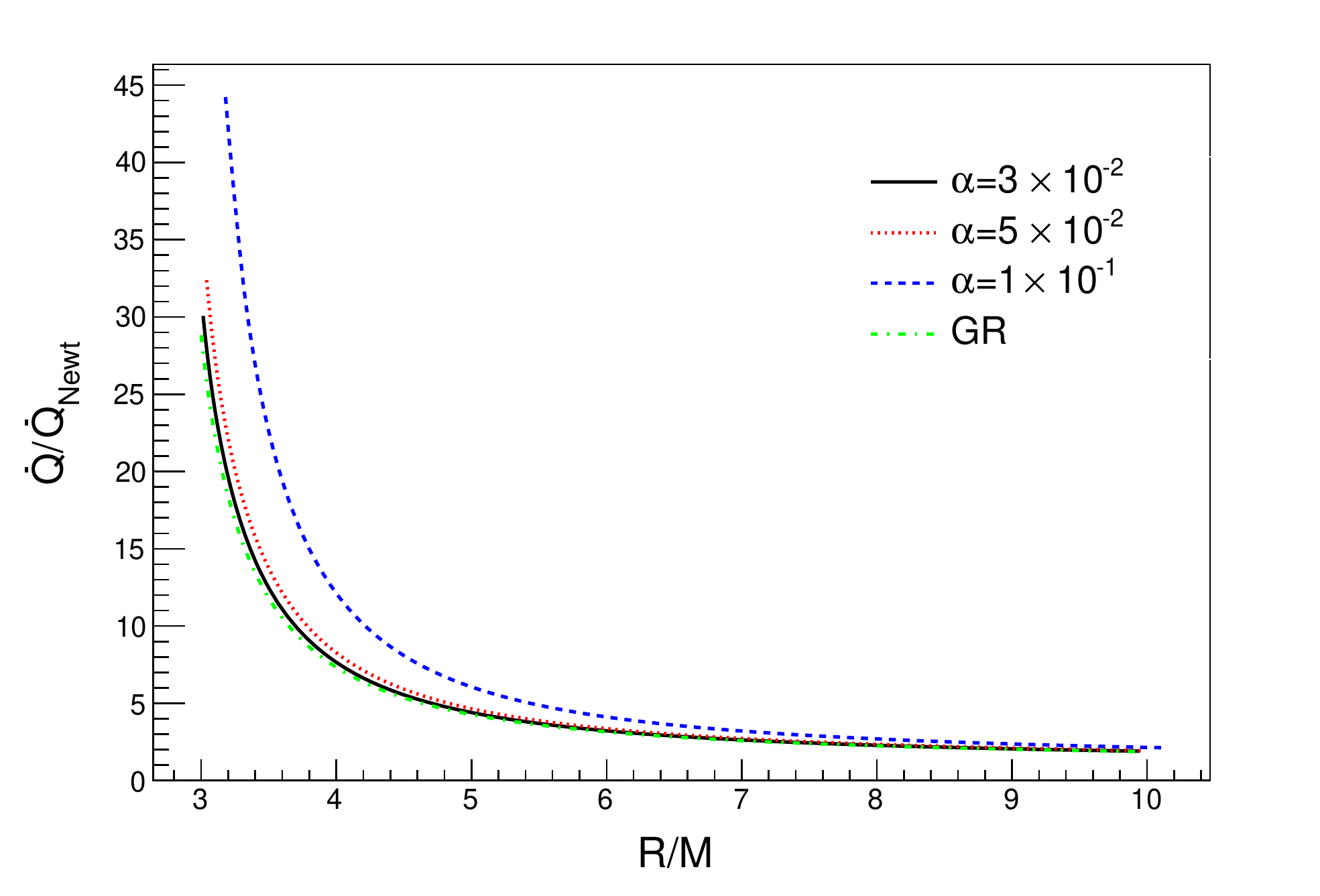}
\caption{Ratio of energy deposition $\dot{Q}$ for the the Sotiriou-Zhau metric to total Newtonian energy deposition $\dot{Q}_{\mathrm{Newt}}$ for different values of $\alpha$. The green curve shows the GR energy deposition for comparison.}
\label{Figmetrica2}
\end{figure}

In what follows, we present three relevant cases that explain how relevant could be the modification to GR in this context. First of all, we take into consideration the Einstein dilaton Gauss-Bonnet gravity. The action is~\cite{Mukherjee:2017fqz}
\begin{equation}
S=\frac{1}{8\pi}\int d^4x\sqrt{-g}\left(\frac{R}{2}-\frac{1}{2}\partial_{\mu}\psi\partial^{\mu}\psi+\alpha\psi L_{GB}\right) \,\ ,
\end{equation}
where $\psi$ is a scalar field and $L_{GB}$ is the Gauss-Bonnet invariant: $L_{GB}=R^2-4R^{\alpha\beta}R_{\alpha\beta}+R^{\alpha\beta\gamma\delta}R_{\alpha\beta\gamma\delta}$. The solution considered is the Sotiriou-Zhau solution(solution in perturbation theory~\citep{Sotiriou:2014pfa}), which lead to results in Fig.~\ref{Figmetrica2}~\cite{Lambiase:2020iul}:
\begin{equation}
ds^2=-f(r)dt^2+h(r)dr^2+r^2d\Omega^2 \,\ ,
\label{metrica2}
\end{equation}
with
\begin{align*}
f(r)&=\left( 1-\frac{2m}{r}\right)\left( 1+\sum_nA_n\bar{a}^n\right) \,\ ;\\
h(r)&=\left( 1-\frac{2m}{r}\right)^{-1}\left( 1+\sum_nB_n\bar{a}^n\right)\,\ .
\end{align*}
where, to the second order:
\begin{align*}
A_1&=B_1=0 \,\ ;\\
A_2&=-\frac{49}{40m^3r}-\frac{49}{20m^2r^2}-\frac{137}{30mr^3}-\frac{7}{15r^4}+\frac{52m}{15r^5}+\frac{40m^2}{3r^6} \,\ ;\\
B_2&=\frac{49}{40m^3r}+\frac{29}{20m^2r^2}+\frac{19}{10mr^3}-\frac{203}{15r^4}-\frac{436m}{15r^5}-\frac{184m^2}{3r^6} \,\ .
\end{align*}
The maximum value taken for $\alpha$, considering the perturbative regime of the solution, shown an increase of the $50\%$ for the maximum amount of energy deposition respect to GR.\\

\begin{figure}
\centering
\includegraphics[scale=0.7]{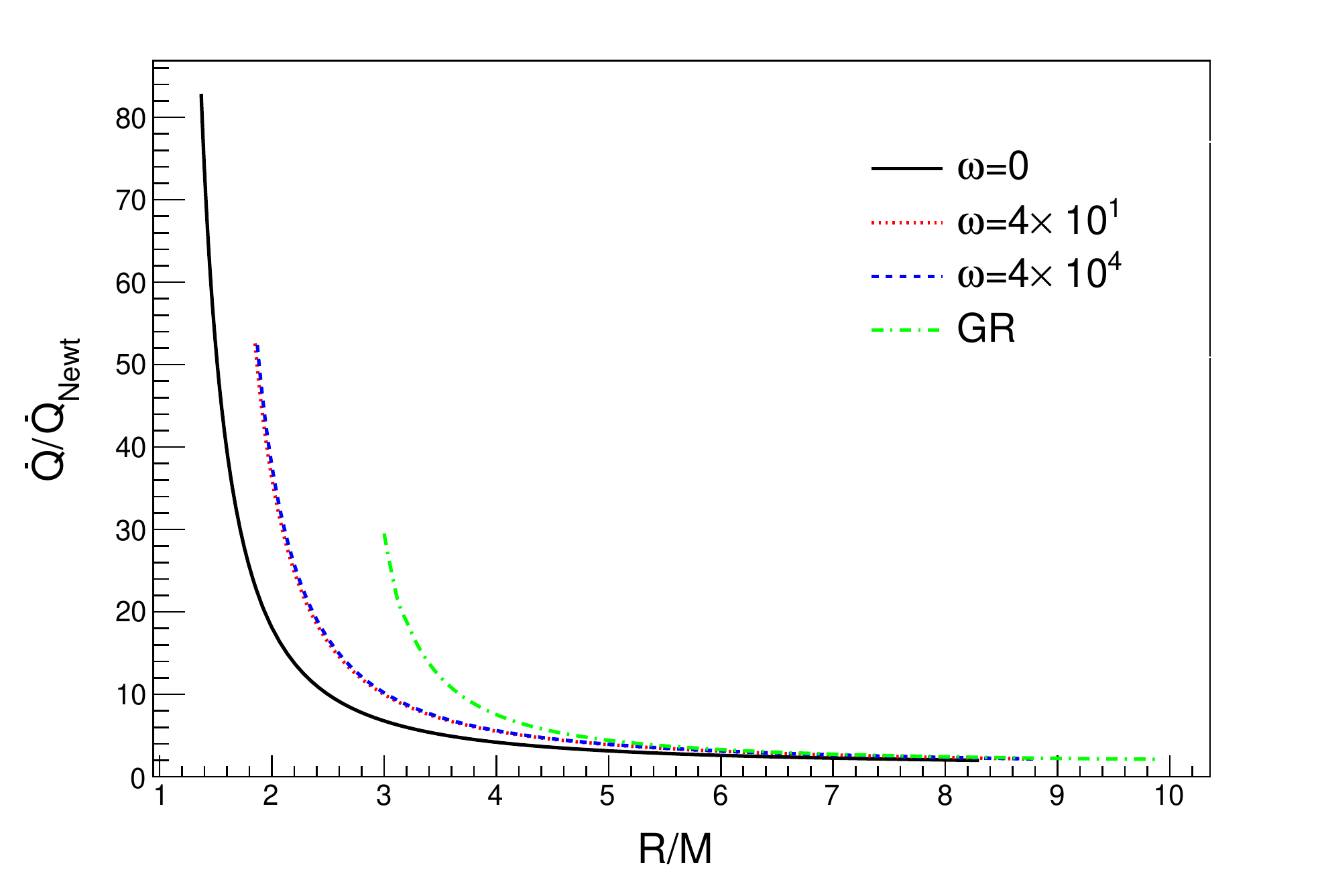}
\caption{Ratio of energy deposition $\dot{Q}$ for metric in Eq.~(\ref{metrica3}) to total Newtonian energy deposition $\dot{Q}_{\mathrm{Newt}}$ for different value of $\omega$. The green curve shows the GR energy deposition for comparison.}
\label{Figmetrica3}
\end{figure}

The second case that we want to analyze is the Brans Dicke theory. It represents a generalization of general relativity, where gravitational effects are in part due to geometry, in part due to a scalar field. The action is~\citep{Brans:1961sx}
\begin{equation}
S=\int d^4x\sqrt{-g}\left[\psi R+\frac{16\pi}{c^4}L-\omega(\psi)\right] \,\ ,
 \end{equation} 
 where $L$ is the Lagrangian density of all the matter, including all non-gravitational field, $\psi$ is a scalar field and $\omega$ is its Lagrangian density.\\
 With this Lagrangian, expressing the line element in the isotropic form, we obtain the solution~\citep{Brans:1961sx}
 \begin{equation}
 ds^2=-e^{2\alpha}dt^2+e^{2\beta}\left[dr^2+r^2d\Omega^2\right] \,\ ,
 \label{metrica3}
 \end{equation}
 where
 \begin{align*}
 \lambda&=\sqrt{(C+1)^2-C(1-\frac{C\omega}{2})} \,\ \\
 e^{2\alpha}&=e^{2\alpha_0}\left[\frac{1-\frac{B}{r}}{1+\frac{B}{r}}\right]^{\frac{2}{\lambda}} \,\ ,\\
 e^{2\beta}&=e^{2\beta_0}\left(1+\frac{B}{r}\right)^4\left[\frac{1-\frac{B}{r}}{1+\frac{B}{r}}\right]^{\frac{2(\lambda-C+1)}{\lambda}} \,\ ,\\
 \psi&=\psi_0\left[\frac{1-\frac{B}{r}}{1+\frac{B}{r}}\right]^{-\frac{C}{\lambda}} \,\ ,
 \end{align*}
 with $\omega$ positive constant and
 \begin{align*}
 \alpha_0&=\beta_0=0 \,\ ,\\
 \psi_0&=\frac{4+2\omega}{3+2\omega} \,\ ,\\
 C&\sim-\frac{1}{2+\omega} \,\ ,\\
 B&\sim\frac{M}{2\sqrt{\psi_0}} \,\ .
 \end{align*}
Using this metric, we obtain the shape for energy deposition in Fig.~\ref{Figmetrica3}. Even with this model we have an enhancement of about $50\%$ respect to the maximum value of $\dot{Q}/\dot{Q_{\mathrm{Newt}}}~30$ in GR. 

\begin{figure}[h!]
\centering
  \includegraphics[scale=0.6]{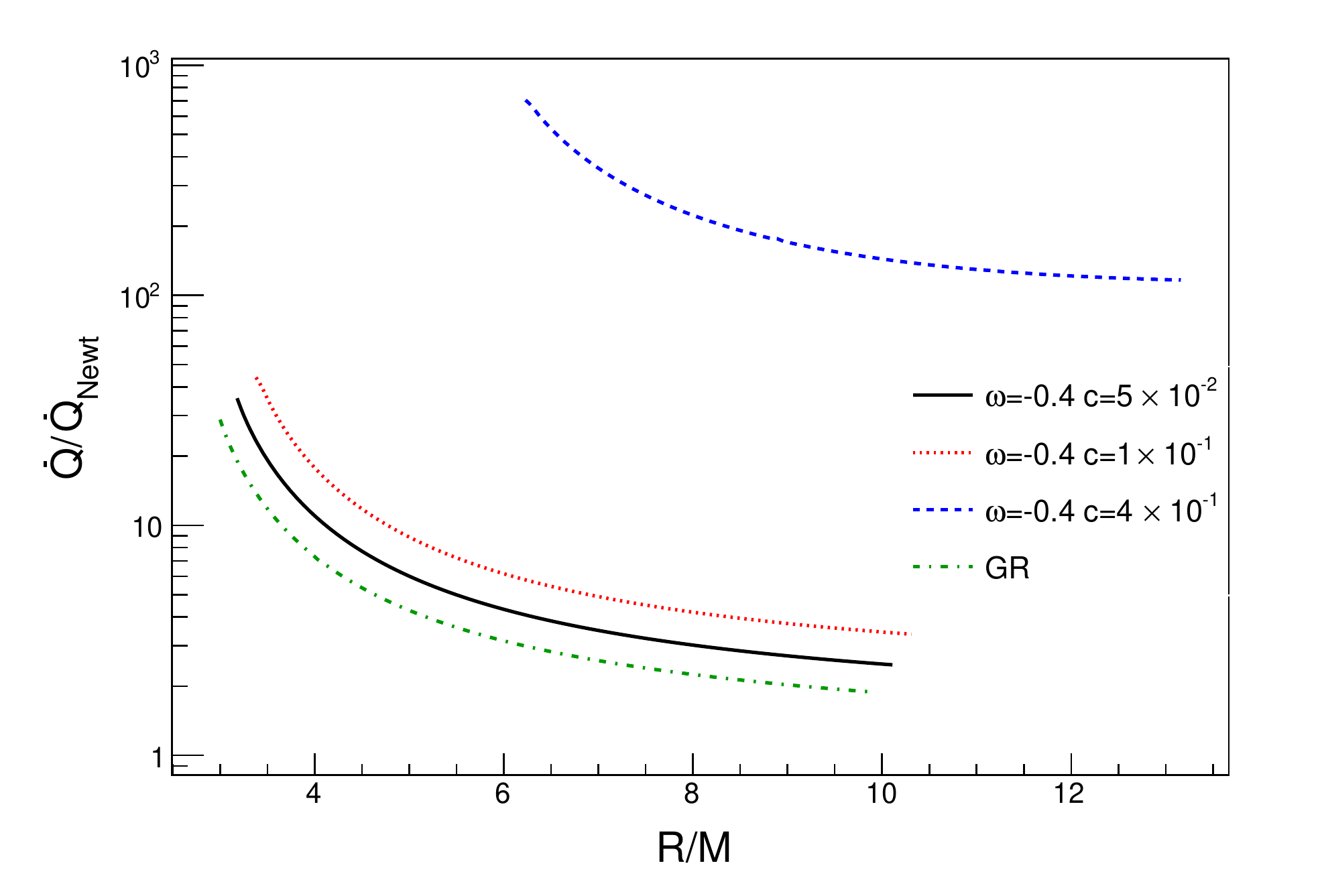}
\caption{Ratio of total energy deposition $\dot{Q}$ for $\omega=-0.4$ to total Newtonian energy deposition $\dot{Q}_{\mathrm{Newt}}$ for three values of the parameter $c$. The green curve shows the GR energy deposition for comparison.}
\label{04}  
\end{figure}

Finally, we discuss the case of a BH surrounded by quintessence field. In this case
\begin{equation}
g_{00}(r)=1-\frac{2M}{r}-\frac{c}{r^{3\omega_q-1}} \,\ ,
\label{metricaq}
\end{equation}
where $c$ is a positive constant and $-1<\omega_q<-1/3$. The quintessence parameter is constrained by the fact that increasing c, the model passes from describing a black hole with an event horizon to representing a naked singularity\\
We chose to show only the case with $\omega=-0.4$, for which we obtain results in Fig.~\ref{04} and an enhancement of a factor $25$ with respect to GR.

\subsection{GRB enhancement}
\begin{figure}[h!]
\centering
  \includegraphics[scale=0.6]{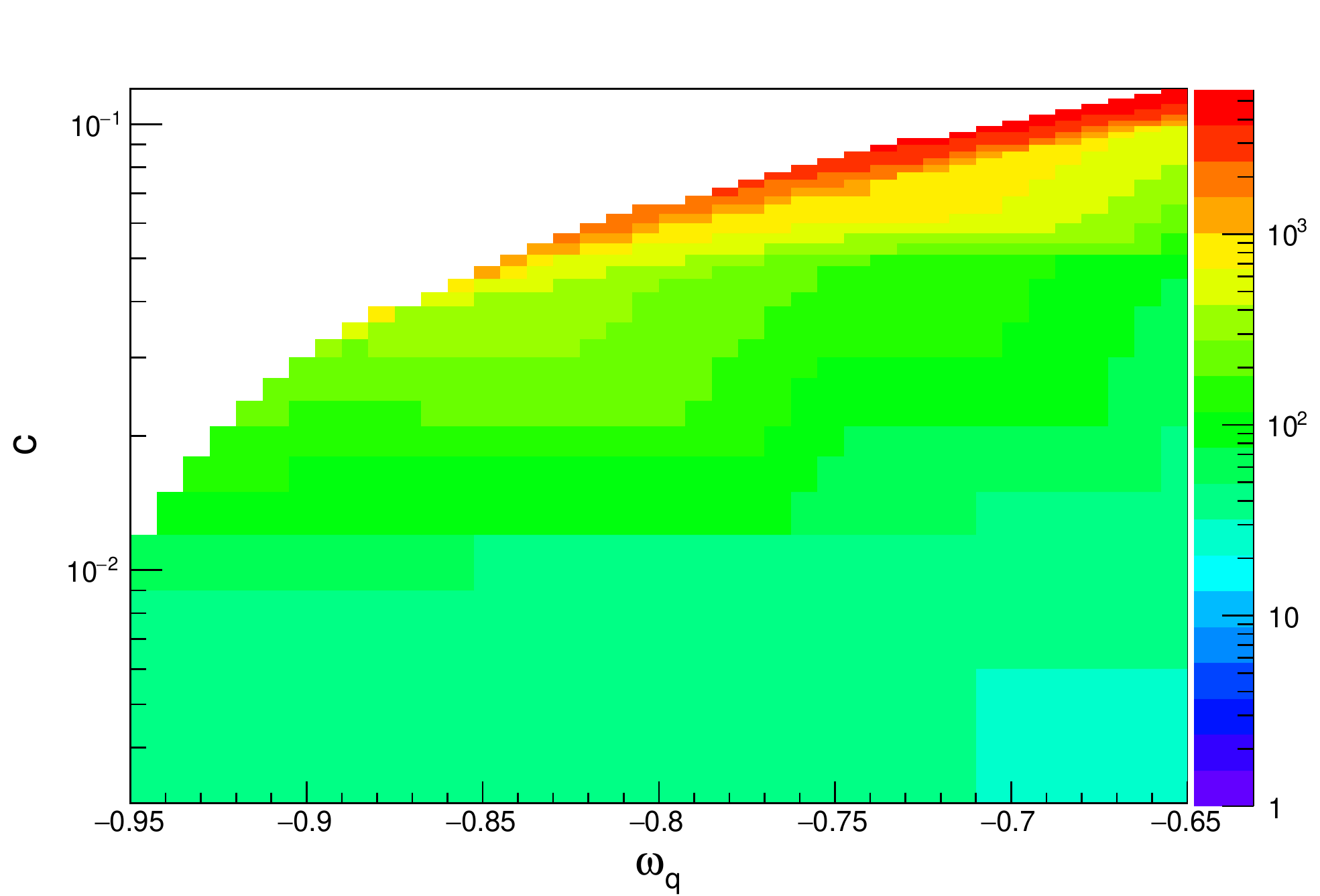}
\caption{Contour plot for $\omega_q\in ]-1,-0.65[$. On the y axis are reported the excluded values of $c$ due to the creation of a naked singularity (white part) and, on the right, the values of $\mathcal{F}(R_{\mathrm{ph}})$. It can be also seen the values of the parameter $c$, for which $\mathcal{F}(R_{\mathrm{ph}}) \sim \mathcal{O}(10^4\textit{-}10^5)$, that are excluded by the energy deposition bounds.}
\label{cont1}       
\end{figure}

\begin{figure}[h!]
\centering
  \includegraphics[scale=0.6]{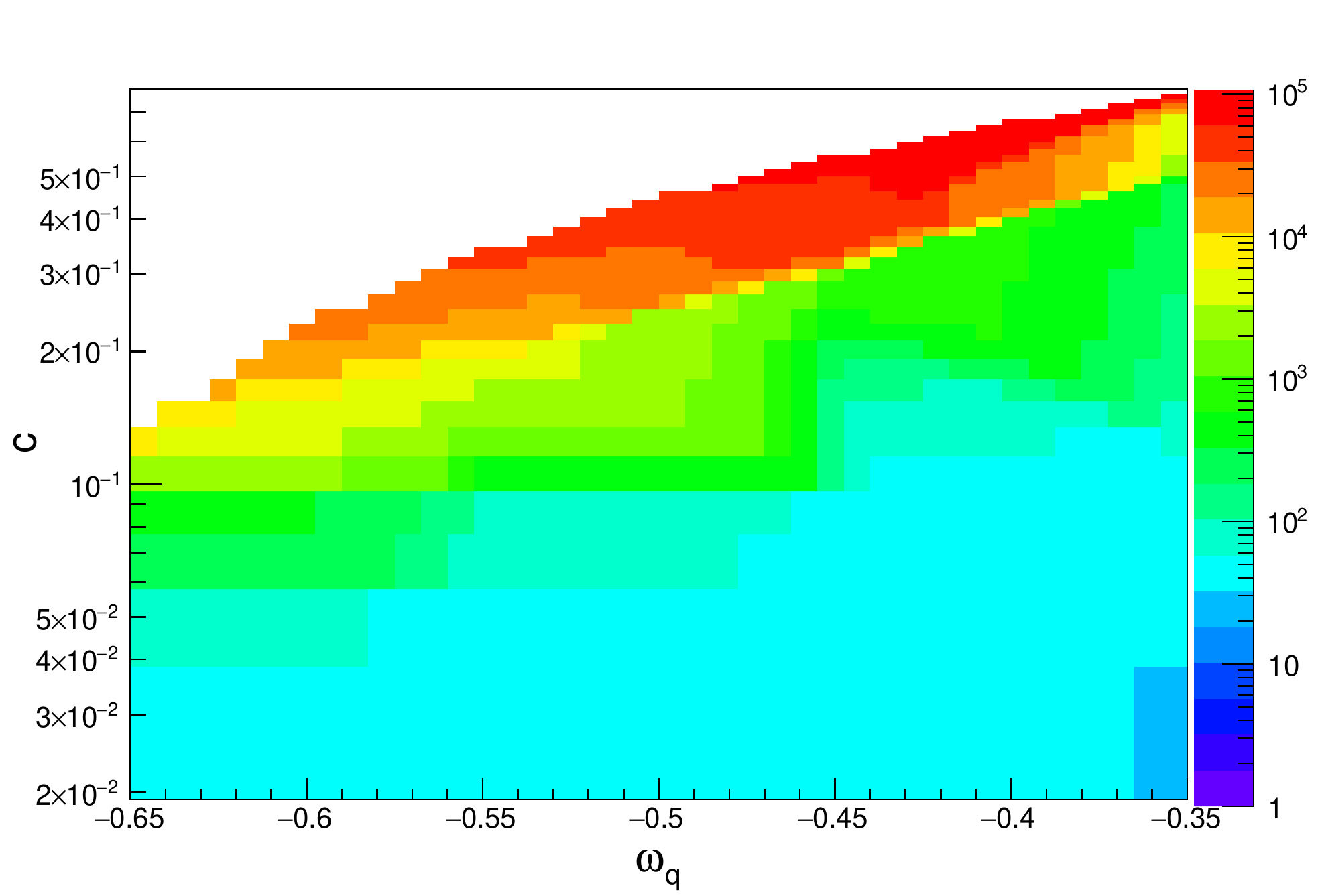}
\caption{Contour plot for $\omega_q\in ]-0.65,-0.35[$. On the y axis are reported the excluded values of $c$ due to the creation of a naked singularity (white part) and, on the right, the values of $\mathcal{F}(R_{\mathrm{ph}})$. It can be also seen the values of the parameter $c$, for which $\mathcal{F}(R_{\mathrm{ph}}) \sim \mathcal{O}(10^4\textit{-}10^5)$, that are excluded by the energy deposition bounds.}
\label{cont2}       
\end{figure}

The above results show that modified gravity provides an enhancement of the neutrino annihilation process as compared to GR. Such an enhancement is relevant for powering GRBs for the model given by a closed neutron stars binary merging system. Neutrino emission happens in the last phase of the merging and the final configuration is a black hole (or neutron star) with an accretion disk. With the developed formalism, we can not describe the disk emission (we are considering a spherical system), so we restrict ourselves to the central BH. Taking into account total energy emitted into neutrinos from the central BH of $\mathcal{O}(10^{52}) ~\mathrm{erg}$, a radius of $R=20~\mathrm{km}$, the maximum possible total energy released in GRB is
\begin{equation}\label{QGR1}
Q_{\mathrm{GR}}\,\,\sim \,\,2.5\times 10^{49}~\mathrm{erg}\,,
\end{equation}
which is too small to explain the short GRB from neutron star merging. Instead, considering the maximum enhancement that modified gravity induces shown in Fig.~\ref{Figmetrica2},\ref{Figmetrica3} and \ref{04}, we have that the total possible energy released in GRB can exceed the maximum energy of $\mathcal{O}(10^{52})~\mathrm{erg}$ (we have to remind that we are considering only the neutrino emission from the central BH and therefore the true emitted energy is larger considering the whole BH+disk configuration).
Moreover, considering that the deposited energy is converted very efficiently to the relativistic jet energy, we infer a constraint on the quintessence model. It is possible to obtain that the maximum allowed value of $\mathcal{F}(R_{ph})$ is $\mathcal{O}(10^4)$ .
The contour plots given in Figs.~\ref{cont1} and \ref{cont2} show the value of $\mathcal{F}(R_{\mathrm{ph}})$ for  allowed value of $\omega_q$ and $c$. Therefore, one can infer, for $\mathcal{F}(R_{\mathrm{ph}}) \sim \mathcal{O}(10^4\textit{-}10^5)$, the values of the parameter $c$ that are not allowed in the considered scenario.


\section{Conclusion}
\label{Conclusion}
In conclusion, we have analyzed the neutrino pair annihilation process $\nu\bar{\nu}\rightarrow e^+ e^-$ near a BH in some modified gravitational theories. We have shown that, owing to a shift of the photosphere radius, there is an enhancement of the deposited energy rate ratio with respect to GR. Such an enhancement could be a relevant mechanism for the generation of GRBs in close neutron star binary merging, for which neutrino pairs annihilation has been proposed as a possible source. Moreover, the released energy may be larger than the Short GRB maximum energy observed of $\mathcal{O}(10^{52})~\mathrm{erg}$ as it happens in the quintessence model (cfr. Ref.~\cite{Lambiase:2020pkc}). In that case, one can constrain the value of $\omega$ and $c$ such that the released energy is inferior to $\mathcal{O}(10^{52})~\mathrm{erg}$.
%
%
The $\nu\bar{\nu}\rightarrow e^+ e^-$ processes are of great importance in astrophysics, as well as in modified gravity, since they lead to considerable differences with respect to GR and its phenomenology. Therefore, the results presented in this proceeding could provide a new astrophysical framework to search for a signature of gravitational theories beyond GR.

\bibliographystyle{ws-procs961x669}
\bibliography{sources1}

\begin{thebibliography}{10}

\bibitem{riess}
A.~G. Riess {\em et~al.}, {Observational evidence from supernovae for an
  accelerating universe and a cosmological constant}, {\em Astron. J.} {\bf
  116}, 1009  (1998).

\bibitem{riess1}
S.~Perlmutter {\em et~al.}, {Measurements of $\Omega$ and $\Lambda$ from 42
  high redshift supernovae}, {\em Astrophys. J.} {\bf 517}, 565  (1999).

\bibitem{riess2}
S.~Cole {\em et~al.}, {The 2dF Galaxy Redshift Survey: Power-spectrum analysis
  of the final dataset and cosmological implications}, {\em Mon. Not. Roy.
  Astron. Soc.} {\bf 362}, 505  (2005).

\bibitem{riess3}
G.~Hinshaw {\em et~al.}, {Three-year Wilkinson Microwave Anisotropy Probe
  (WMAP) observations: temperature analysis}, {\em Astrophys. J. Suppl.} {\bf
  170}, p. 288  (2007).

\bibitem{riess4}
S.~M. Carroll, {The Cosmological constant}, {\em Living Rev. Rel.} {\bf 4},
  p.~1  (2001).

\bibitem{riess5}
V.~Sahni and A.~A. Starobinsky, {The Case for a positive cosmological Lambda
  term}, {\em Int. J. Mod. Phys. D} {\bf 9}, 373  (2000).

\bibitem{starobinski}
A.~A. Starobinsky, {A New Type of Isotropic Cosmological Models Without
  Singularity}, {\em Adv. Ser. Astrophys. Cosmol.} {\bf 3}, 130  (1987).

\bibitem{starobinski1}
A.~A. Starobinsky, {The Perturbation Spectrum Evolving from a Nonsingular
  Initially De-Sitter Cosmology and the Microwave Background Anisotropy}, {\em
  Sov. Astron. Lett.} {\bf 9}, p. 302  (1983).

\bibitem{cosmo2}
H.~Oyaizu, M.~Lima and W.~Hu, {Nonlinear evolution of $f(R)$ cosmologies. II.
  Power spectrum}, {\em Phys. Rev. D} {\bf 78}, p. 123524 (Dec 2008).

\bibitem{cosmo3}
L.~Pogosian and A.~Silvestri, {The pattern of growth in viable f(R)
  cosmologies}, {\em Phys. Rev. D} {\bf 77}, p. 023503  (2008), [Erratum:
  Phys.Rev.D 81, 049901 (2010)].

\bibitem{cosmo4}
I.~Sawicki and W.~Hu, {Stability of cosmological solutions in $f(R)$ models of
  gravity}, {\em Phys. Rev. D} {\bf 75}, p. 127502 (Jun 2007).

\bibitem{cosmo5}
B.~Li and J.~D. Barrow, {Cosmology of $f(R)$ gravity in the metric variational
  approach}, {\em Phys. Rev. D} {\bf 75}, p. 084010 (Apr 2007).

\bibitem{cosmo6}
T.~Clifton, Higher powers in gravitation, {\em Phys. Rev. D} {\bf 78}, p.
  083501 (Oct 2008).

\bibitem{cosmo7}
T.~Clifton and J.~D. Barrow, The power of general relativity, {\em Phys. Rev.
  D} {\bf 72}, p. 103005 (Nov 2005).

\bibitem{cosmo8}
S.~Capozziello and G.~Lambiase, {Higher order corrections to the effective
  gravitational action from Noether symmetry approach}, {\em Gen. Rel. Grav.}
  {\bf 32}, 295  (2000).

\bibitem{cosmo9}
S.~Capozziello and G.~Lambiase, {Nonminimal derivative coupling and the
  recovering of cosmological constant}, {\em Gen. Rel. Grav.} {\bf 31}, 1005
  (1999).

\bibitem{cosmo10}
S.~Capozziello, G.~Lambiase and H.~Schmidt, {Nonminimal derivative couplings
  and inflation in generalized theories of gravity}, {\em Annalen Phys.} {\bf
  9}, 39  (2000).

\bibitem{cosmo11}
S.~Nojiri and S.~D. Odintsov, Inhomogeneous equation of state of the universe:
  Phantom era, future singularity, and crossing the phantom barrier, {\em Phys.
  Rev. D} {\bf 72}, p. 023003 (Jul 2005).

\bibitem{cosmo12}
S.~Capozziello, V.~F. Cardone, E.~Elizalde, S.~Nojiri and S.~D. Odintsov,
  Observational constraints on dark energy with generalized equations of state,
  {\em Phys. Rev. D} {\bf 73}, p. 043512 (Feb 2006).

\bibitem{cosmo13}
I.~Brevik, E.~Elizalde, S.~Nojiri and S.~D. Odintsov, Viscous little rip
  cosmology, {\em Phys. Rev. D} {\bf 84}, p. 103508 (Nov 2011).

\bibitem{cosmo14}
S.~Nojiri and S.~D. Odintsov, {The New form of the equation of state for dark
  energy fluid and accelerating universe}, {\em Phys. Lett. B} {\bf 639}, 144
  (2006).

\bibitem{cosmo15}
S.~Nojiri and S.~D. Odintsov, {Non-singular modified gravity unifying inflation
  with late-time acceleration and universality of viscous ratio bound in F(R)
  theory}, {\em Prog. Theor. Phys. Suppl.} {\bf 190}, 155  (2011).

\bibitem{cosmo17}
G.~Lambiase, {Thermal leptogenesis in $f(R)$ cosmology}, {\em Phys. Rev. D}
  {\bf 90}, p. 064050 (Sep 2014).

\bibitem{cosmo18}
G.~Lambiase, S.~Mohanty and A.~R. Prasanna, {Neutrino coupling to cosmological
  background: A review on gravitational Baryo/Leptogenesis}, {\em Int. J. Mod.
  Phys. D} {\bf 22}, p. 1330030  (2013).

\bibitem{cosmo19}
G.~Lambiase and G.~Scarpetta, {Baryogenesis in $f(R)$ theories of gravity},
  {\em Phys. Rev. D} {\bf 74}, p. 087504 (Oct 2006).

\bibitem{cosmo20}
B.~Jain, V.~Vikram and J.~Sakstein, {Astrophysical Tests of Modified Gravity:
  Constraints from Distance Indicators in the Nearby Universe}, {\em Astrophys.
  J.} {\bf 779}, p.~39  (2013).

\bibitem{cosmo21}
L.~Lombriser, A.~Slosar, U.~Seljak and W.~Hu, {Constraints on f(R) gravity from
  probing the large-scale structure}, {\em Phys. Rev. D} {\bf 85}, p. 124038
  (2012).

\bibitem{cosmo22}
S.~Ferraro, F.~Schmidt and W.~Hu, {Cluster abundance in $f(R)$ gravity models},
  {\em Phys. Rev. D} {\bf 83}, p. 063503 (Mar 2011).

\bibitem{cosmo23}
L.~Lombriser, F.~Schmidt, T.~Baldauf, R.~Mandelbaum, U.~c.~v. Seljak and R.~E.
  Smith, Cluster density profiles as a test of modified gravity, {\em Phys.
  Rev. D} {\bf 85}, p. 102001 (May 2012).

\bibitem{cosmo24}
F.~Schmidt, A.~Vikhlinin and W.~Hu, {Cluster constraints on $f(R)$ gravity},
  {\em Phys. Rev. D} {\bf 80}, p. 083505 (Oct 2009).

\bibitem{cosmo25}
H.~Motohashi, A.~A. Starobinsky and J.~Yokoyama, {Cosmology Based on f(R)
  Gravity Admits 1~eV Sterile Neutrinos}, {\em Phys. Rev. Lett.} {\bf 110}, p.
  121302  (2013).

\bibitem{Buoninfante:2021qrv}
L.~Buoninfante, G.~Lambiase and L.~Petruzziello, {Quantum interference in
  external gravitational fields beyond General Relativity} (4 2021).

\bibitem{Capolupo:2021blb}
A.~Capolupo, G.~Lambiase, A.~Stabile and A.~Stabile, {Virial theorem in scalar
  tensor fourth order gravity}, {\em Eur. Phys. J. C} {\bf 81}, p. 650  (2021).

\bibitem{Bittencourt:2020lgu}
V.~A. S.~V. Bittencourt, M.~Blasone, F.~Illuminati, G.~Lambiase, G.~G. Luciano
  and L.~Petruzziello, {Quantum nonlocality in extended theories of gravity},
  {\em Phys. Rev. D} {\bf 103}, p. 044051  (2021).

\bibitem{Lambiase:2020vul}
G.~Lambiase, M.~Sakellariadou and A.~Stabile, {Constraints on extended gravity
  models through gravitational wave emission}, {\em JCAP} {\bf 03}, p. 014
  (2021).

\bibitem{Bernal:2020ywq}
N.~Bernal, A.~Ghoshal, F.~Hajkarim and G.~Lambiase, {Primordial Gravitational
  Wave Signals in Modified Cosmologies}, {\em JCAP} {\bf 11}, p. 051  (2020).

\bibitem{Tino:2020nla}
G.~Tino, L.~Cacciapuoti, S.~Capozziello, G.~Lambiase and F.~Sorrentino,
  {Precision Gravity Tests and the Einstein Equivalence Principle}, {\em Prog.
  Part. Nucl. Phys.} {\bf 112}, p. 103772  (2020).

\bibitem{1999A&A...344..573R}
M.~{Ruffert} and H.~T. {Janka}, {Gamma-ray bursts from accreting black holes in
  neutron star mergers}, {\em Astron. Astrophys.} {\bf 344}, 573 (April 1999).

\bibitem{Popham_1999}
R.~Popham, S.~E. Woosley and C.~Fryer, Hyperaccreting black holes and gamma-ray
  bursts, {\em The Astrophysical Journal} {\bf 518}, 356 (jun 1999).

\bibitem{Di_Matteo_2002}
T.~D. Matteo, R.~Perna and R.~Narayan, Neutrino trapping and accretion models
  for gamma-ray bursts, {\em The Astrophysical Journal} {\bf 579}, 706 (nov
  2002).

\bibitem{Fujibayashi_2017}
S.~Fujibayashi, Y.~Sekiguchi, K.~Kiuchi and M.~Shibata, Properties of
  neutrino-driven ejecta from the remnant of a binary neutron star merger: Pure
  radiation hydrodynamics case, {\em The Astrophysical Journal} {\bf 846}, p.
  114 (sep 2017).

\bibitem{Just:2015dba}
O.~Just, M.~Obergaulinger, H.~T. Janka, A.~Bauswein and N.~Schwarz,
  {Neutron-star merger ejecta as obstacles to neutrino-powered jets of
  gamma-ray bursts}, {\em Astrophys. J. Lett.} {\bf 816}, p. L30  (2016).

\bibitem{PhysRevD.98.063007}
F.~Foucart, M.~D. Duez, L.~E. Kidder, R.~Nguyen, H.~P. Pfeiffer and M.~A.
  Scheel, Evaluating radiation transport errors in merger simulations using a
  monte carlo algorithm, {\em Phys. Rev. D} {\bf 98}, p. 063007 (Sep 2018).

\bibitem{Salmonson:1999es}
J.~D. Salmonson and J.~R. Wilson, {General relativistic augmentation of
  neutrino pair annihilation energy deposition near neutron stars}, {\em
  Astrophys. J.} {\bf 517}, 859  (1999).

\bibitem{Prasanna:2001ie}
A.~Prasanna and S.~Goswami, {Energy deposition due to neutrino pair
  annihilation near rotating neutron stars}, {\em Phys. Lett. B} {\bf 526}, 27
  (2002).

\bibitem{Lambiase:2020iul}
G.~Lambiase and L.~Mastrototaro, {Effects of modified theories of gravity on
  neutrino pair annihilation energy deposition near neutron stars}, {\em ApJ.}
  {\bf 904}, 1  (2020).

\bibitem{Co86}
J.~Cooperstein, L.~van~den Horn and E.~A. Baron, {Neutrino flows in collapsing
  stars: a two-fluid model}, {\em ApJ} {\bf 309}, p. 653  (1986).

\bibitem{Co87}
J.~Cooperstein, L.~J. van~den Horn and E.~A. Baron, {Neutrino Pair Energy
  Deposition in Supernovae}, {\em ApJ} {\bf 321}, p. L129  (1987).

\bibitem{Salmonson:2001tz}
J.~D. Salmonson and J.~R. Wilson, {Neutrino annihilation between binary neutron
  stars}, {\em Astrophys. J.} {\bf 561}, 950  (2001).

\bibitem{Bhattacharyya:2009nm}
R.~Mallick, A.~Bhattacharyya, S.~K. Ghosh and S.~Raha, {General Relativistic
  effect on the energy deposition rate for neutrino pair annihilation above the
  equatorial plane along the symmetry axis near a rotating neutron star}, {\em
  Int. J. Mod. Phys. E} {\bf 22}, p. 1350008  (2013).

\bibitem{Goodman:1986we}
J.~Goodman, A.~Dar and S.~Nussinov, {Neutrino Annihilation in type II
  Supernovae}, {\em Astrophys. J. Lett.} {\bf 314}, L7  (1987).

\bibitem{Mukherjee:2017fqz}
S.~Mukherjee and S.~Chakraborty, {Horndeski theories confront the Gravity Probe
  B experiment}, {\em Phys. Rev. D} {\bf 97}, p. 124007  (2018).

\bibitem{Sotiriou:2014pfa}
T.~P. Sotiriou and S.-Y. Zhou, {Black hole hair in generalized scalar-tensor
  gravity: An explicit example}, {\em Phys. Rev. D} {\bf 90}, p. 124063
  (2014).

\bibitem{Brans:1961sx}
C.~Brans and R.~Dicke, {Mach's principle and a relativistic theory of
  gravitation}, {\em Phys. Rev.} {\bf 124}, 925  (1961).

\bibitem{Lambiase:2020pkc}
G.~Lambiase and L.~Mastrototaro, {Neutrino pair annihilation ($\nu{\bar \nu}\to
  e^-e^+$) in the presence of quintessence surrounding a black hole} (12 2020).

\end{thebibliography}

\end{document}